\documentclass[3p,english,12pt,times,sort&compress,nopreprintline]{elsarticle}
\usepackage{fullpage}

\usepackage[utf8]{inputenc}
\usepackage[T1]{fontenc}
\usepackage{babel}
\usepackage{array}
\usepackage{multirow}
\usepackage{amsmath}
\usepackage{amsthm}
\usepackage{amssymb}
\usepackage{nicefrac}
\usepackage{graphicx}
\usepackage[unicode=true]{hyperref}
\usepackage{xcolor}
\usepackage{textgreek}
\usepackage{comment}
\usepackage[title]{appendix}
\usepackage{subfig}
\usepackage{longtable}
\usepackage{multirow}
\usepackage{booktabs}
\usepackage{caption}
\usepackage{array}
\usepackage{calc}
\usepackage{etoolbox}

\hypersetup{
  colorlinks=true,
  linkcolor=blue,
  filecolor=magenta,
  urlcolor=cyan,
}

\makeatletter
\newcommand{\orcidID}[1]{\href{https://orcid.org/#1}{\includegraphics[width=10pt]{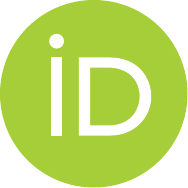}}}
\def\autocite#1\citep{#1}
\theoremstyle{plain}

\newtheorem{theorem}{\protect\theoremname}
\newtheorem*{theorem*}{\protect\theoremname}
\theoremstyle{plain}

\newtheorem{definition}[theorem]{\protect\definitionname}
\newtheorem*{definition*}{\protect\definitionname}
\theoremstyle{plain}

\newtheorem{remark}[theorem]{\protect\remarkname}
\newtheorem*{remark*}{\protect\remarkname}
\theoremstyle{plain}

\newtheorem{lemma}[theorem]{\protect\lemmaname}
\newtheorem*{lemma*}{\protect\lemmaname}
\theoremstyle{plain}
\newtheorem{corollary}[theorem]{\protect\corollaryname}
\newtheorem*{corollary*}{\protect\corollaryname}
\makeatother

\providecommand{\definitionname}{Definition}
\providecommand{\lemmaname}{Lemma}
\providecommand{\remarkname}{Remark}
\providecommand{\theoremname}{Theorem}
\providecommand{\corollaryname}{Corollary}

\usepackage{aliascnt}
\makeatletter
\let\corollary\@undefined
\let\endcorollary\@undefined
\let\lemma\@undefined
\let\endlemma\@undefined
\let\definition\@undefined
\let\enddefinition\@undefined
\makeatother
\theoremstyle{plain}
\newaliascnt{lemma}{theorem}
\newtheorem{lemma}[lemma]{Lemma}
\aliascntresetthe{lemma}

\newaliascnt{corollary}{theorem}
\newtheorem{corollary}[corollary]{Corollary}
\aliascntresetthe{corollary}

\newaliascnt{definition}{theorem}
\newtheorem{definition}[definition]{Definition}
\aliascntresetthe{definition}

\makeatletter
\newcommand\Autoref[1]{\@first@ref#1,@}
\def\@throw@dot#1.#2@{#1}
\def\@set@refname#1{
    \edef\@tmp{\getrefbykeydefault{#1}{anchor}{}}%
    \xdef\@tmp{\expandafter\@throw@dot\@tmp.@}%
    \ltx@IfUndefined{\@tmp autorefnameplural}%
         {\def\@refname{\@nameuse{\@tmp autorefname}s}}%
         {\def\@refname{\@nameuse{\@tmp autorefnameplural}}}%
}
\def\@first@ref#1,#2{%
  \ifx#2@\autoref{#1}\let\@nextref\@gobble
  \else%
    \@set@refname{#1}
    \@refname~\ref{#1}
    \let\@nextref\@next@ref
  \fi%
  \@nextref#2%
}
\def\@next@ref#1,#2{%
   \ifx#2@ and~\ref{#1}\let\@nextref\@gobble
   \else, \ref{#1}
   \fi%
   \@nextref#2%
}
\makeatother

\makeatletter
\def\maxwidth{\ifdim\Gin@nat@width>\linewidth\linewidth\else\Gin@nat@width\fi}
\def\maxheight{\ifdim\Gin@nat@height>\textheight\textheight\else\Gin@nat@height\fi}
\makeatother
\setkeys{Gin}{width=\maxwidth,height=\maxheight,keepaspectratio}

\begin{document}

\begin{frontmatter}

\title{SCAU: Modeling spectral causality for multivariate time series with applications to electroencephalograms}

\author[mek]{Marco Antonio Pinto Orellana \orcidID{0000-0001-6495-1305}}
\corref{cor}
\author[mek]{Peyman Mirtaheri \orcidID{0000-0002-7664-5513}}
\author[it,simulamet]{Hugo L. Hammer \orcidID{0000-0001-9429-7148}}
\author[kaust]{Hernando Ombao \orcidID{0000-0001-7020-8091}}

\cortext[cor]{Corresponding author}

\address[mek]{Department of Mechanical, Electronics and Chemical Engineering. Oslo Metropolitan University.}
\address[it]{Department of Information Technology. Oslo Metropolitan University.}
\address[simulamet]{Department of Holistic Systems, Simula Metropolitan Center for Digital Engineering.}
\address[kaust]{Biostatistics research group. King Abdullah University of Sciences and Technology.}

\begin{abstract}
Electroencephalograms (EEG) are noninvasive measurement signals of electrical neuronal activity in the brain. One of the current major statistical challenges is formally measuring functional dependency between those complex signals. This paper, proposes the spectral causality model (SCAU), a robust linear model, under a causality paradigm, to reflect inter- and intra-frequency modulation effects that cannot be identifiable using other methods. SCAU inference is conducted with three main steps: (a) signal decomposition into frequency bins, (b) intermediate spectral band mapping, and (c) dependency modeling through frequency-specific autoregressive models (VAR). We apply SCAU to study complex dependencies during visual and lexical fluency tasks (word generation and visual fixation) in 26 participants' EEGs. We compared the connectivity networks estimated using SCAU with respect to a VAR model. SCAU networks show a clear contrast for both stimuli while the magnitude links also denoted a low variance in comparison with the VAR networks. Furthermore, SCAU dependency connections not only were consistent with findings in the neuroscience literature, but it also provided further evidence on the directionality of the spatio-spectral dependencies such as the delta-originated and theta-induced links in the fronto-temporal brain network.
\end{abstract}


\end{frontmatter}

\section{Introduction}

\subsection{Modulation in biological settings}

Electroencephalograms (EEGs) are multivariate time series recorded from many points in space at the scalp as result of the electrical activity generated by numerous and synchronized group of pyramidal neurons. EEGs are often analyzed through their spectral properties, given the association between physiological conditions and specific frequency intervals. The most comprehensive analysis method relies on dividing the complete spectrum into five major ranges: delta (0-4Hz), theta (4-8Hz), alpha (8-12Hz), beta (12-30Hz), and gamma rhythms ($> 30$Hz) \citep[ pp.~47]{AtlasEEGSeizure-Khalil-2006}. Note that the exact borders can vary according to the researcher. Each frequency bin on this division can also be related to some physical, healthy or abnormal, condition: delta waves often observed during sleep, theta rhythms common in mental imagery; alpha waves visible in the occipital region during resting states with closed eyes (Berger effect); and beta waves which are related to alertness and can be affected by the consumption of certain types of medication \citetext{\citealp[ pp.~28-34]{HandbookEEGInterpretation-Tatum-2008}; \citealp[ pp.~47-48]{AtlasEEGSeizure-Khalil-2006}}.

Modulation, in a general sense, is understood as the phenomena where an external or internal signal drives (or modulates) the parameters of another signal (carrier signal) \citep[ pp.~297]{SigSystemsBiomed-Devasahayam-2000}. This abstraction, typical in communication systems, has often been described in neuroscience literature to model some neurological processes. Even more, pulse code modulation is the foundational model to describe the effect of action potential firing spikes sequences on the neurons' axon hillocks \citep[ pp.~299]{SigSystemsBiomed-Devasahayam-2013}. Our contribution, in this sense, is a model that can adequately represent these phenomena in a statistical setting.

In EEG, implicit modulation appears in different scenarios. The Berger effect described that signal representing the eye closeness status modulates the EEG generating a specific type of signals known as posterior dominant rhythms with frequencies in the alpha region. If this signal only appears in one hemisphere or denotes a frequency in the theta interval, it is a marker of seizures or other encephalopathies \citep[ pp.~48]{AtlasEEGSeizure-Khalil-2006}. Nozaradan et al.~also investigated the feasibility of frequency-controlled music as stimuli as a modulator of EEG signals using a frequency tagging technique. Their discoveries described a ``nonlinear transformation of the sound envelope'' observed in the EEG signals \citep{SteadystateEvokedPotentials-Nozaradan-2012}. Orekhova et al.~also reported that observed gamma waves could have their main frequencies modulated by the speed of visual stimuli (although the modulation parameters can be affected by age) \citep{FreqGammaOsc-Orekhova-2015}. \\Furthermore, Albada et al.~also show some interesting linear relationships between beta and alpha bands, observing a linear relationship between the peaks' magnitudes \citep{EEGSpectral-Albada-2013}. And also, Sato et al.~denoted that interactions on specific brain regions close to the occipital lobe can manifest gamma-gamma modulation effects during face processing \citep{BiDirElec-Sato-2017}. Despite their biological relevance, straightforward linear models present some limitations for capturing the dynamics under cross-frequency modulation. As an alternative, we propose a model that can address the challenges of modeling such modulated signals under data transformations using a Granger causality framework.

\subsection{Granger causality and under vector autoregressive models}

Granger causality, or time-causality, in time series is often described in the context of a vector autoregressive (VAR) process. A multivariate time series is said to follow a vector autoregressive model of order $p$ if it explicitly represents the current value of a system as a linear combination of the previous $p$ time points \citep[pp.~272-288]{TimeSeries-Shumway-2017}:%
\begin{equation}%
{
x\left(n\right)=\sum_{\ell=1}^{p}\Phi_{\ell}x\left(n-\ell\right)+\varepsilon\left(n\right)
}\label{eq:var-model}%
\end{equation}%
where $\varepsilon\left(n\right)$ is a $p\times1$ i.i.d random vector with covariance matrix $\Sigma$: $\varepsilon\left(n\right)\sim N_{p}\left(0,\Sigma\right)$, and $\Phi_{\ell}$ is a $k\times k$ transition matrix that ``expresses the dependency'' of $x\left(n - \ell\right)$ on $x\left(n\right)$.

Granger causality (GC) establishes a relationship of cause-effect between components (or channels) of a multivariate time series. Under this framework, a time series $x\left(n\right)$ is said to cause another time series $z\left(n\right)$ in Granger's sense (or $x\left(n\right)$ is Granger-causal for $z\left(n\right)$ ) when the mean squared error (MSE) of the linear $h$-step forecast of $z\left(n\right)$ is reduced if $x\left(\tau\right),\tau<t$ is included as a covariate to predict $z\left(n\right)$ \citep[ pp.~41-42]{NewIntroductionMultiple-Lutkepohl-2005}. Thus, Granger causality under a VAR model is established as follows. Given that $x_{i}\left(n\right)$ Granger-causes $x_{j}\left(n\right)$ if there is a coefficient $\phi_{i,j}^{\ell}\ne0\, \vert \, j=1,\ldots,N,\ell=1,\ldots,p$ \citep[pp.~45, corollary 2.2.1]{NewIntroductionMultiple-Lutkepohl-2005}, then GC $x_{i}\mapsto x_{j}$ can be assessed by testing the statistical significance of the estimate $\hat{\phi}_{i,j}^{\ell}$. In general, VAR parameters are often estimated with multivariate least squares as described in \citep[ pp.~69-82]{NewIntroductionMultiple-Lutkepohl-2005}. VAR models  effectively detect linear dependencies when there is no exact multicollinearity among the least square covariates.

To the best of our knowledge, no study attempts to provide a causality model to represent interactions at a spatial-frequency, or channel-frequency, level. Onton et al.~introduced the closest proposal that relates coherence between spectrum intervals using independent component analysis (ICA) in a spatial-frequency framework \citep{HighFreqBroadbandModulation-Onton-2009}. However, the proposed statistical independent component cannot offer direct biological interpretability, and the method was not extended to a causality inference context.

Under modulation phenomena, straightforward VAR models could not capture cross-frequency dependencies, as will be discussed in the following sections. This paper provides a framework for the analysis of causality to describe the dynamics between the frequency components that composed every channel in EEG recordings. Our proposal is based on an extension of VAR models to analyze time series with relevant spectral information that is assumed to be modulated. Even though our analysis is focused on EEG, the spectral causality framework can also be easily extended to any other type of time series where the frequency components perform a major and interpretable role.

\section{Spectral causality framework}

\begin{figure}
\hypertarget{fig:graphical-summary}{%
\centering
\includegraphics[width=0.95\textwidth,height=\textheight]{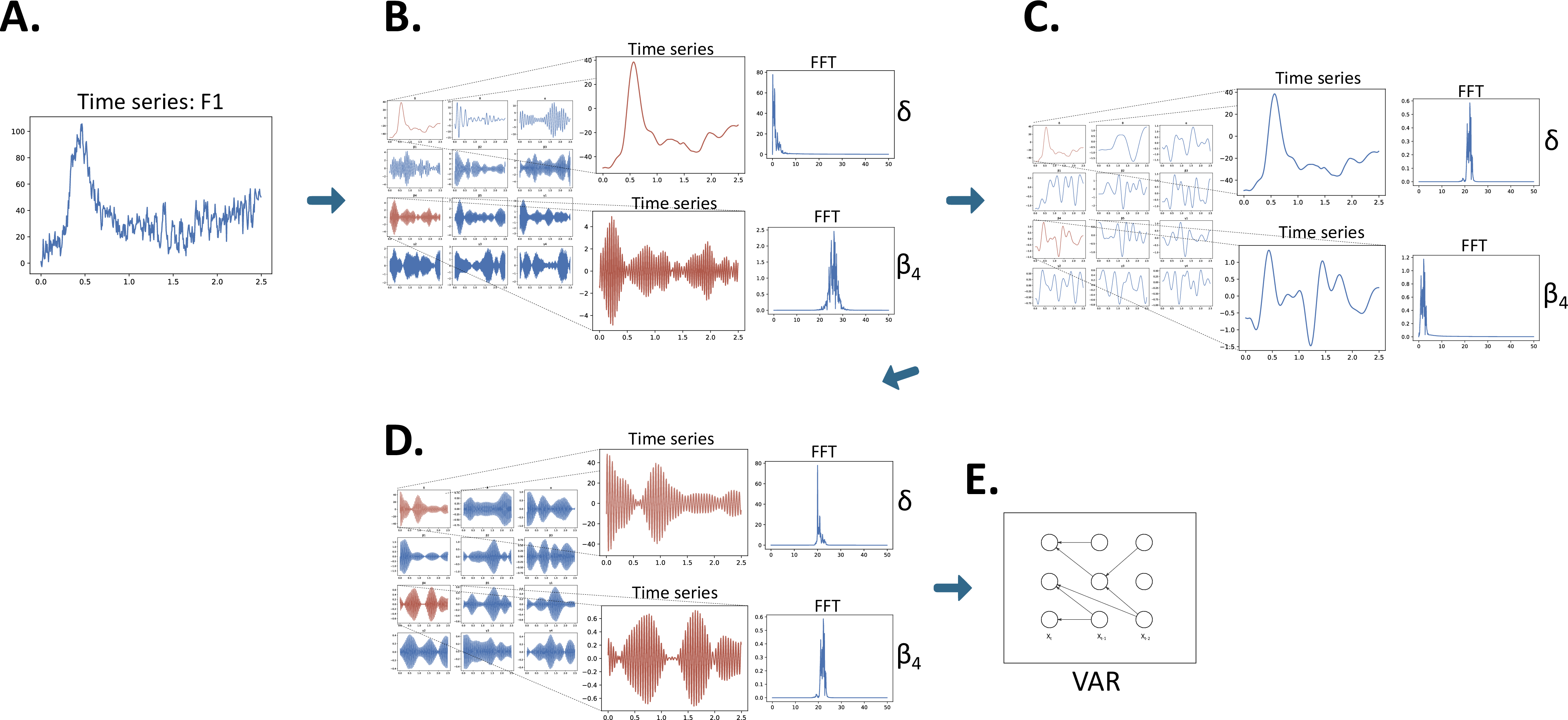}
\caption{Graphical summary of the spectral causality framework: a raw time series (A) is decomposed into frequency intervals; (B) the components are mapped to a lower frequency region with a phase of 180 degrees; (C) to restore the original phase and minimize estimation issues, the channels are mapped to an intermediate frequency; and finally, (D) the dependency and causality are evaluated through a vector autoregressive (VAR) model.}\label{fig:graphical-summary}
}
\end{figure}

The general notion of causality in multivariate time series confirms whether or not fluctuations in one component ``cause'' changes in another. This provides essential information about how apriori knowledge of the historical value of one component can help reduce the forecast error of another. However, it does not indicate the contributions of the different waveforms (of various frequencies) to the causality relationship. The goal of this paper is to address this severe limitation by spectral causality. Our proposed approach aims to provide a causality framework that captures the dependency among signal frequency components. In essence, the method consists of: (1.) decompose the signal into frequency bands of research interest by one-sided linear filtering; (2.) apply a frequency transformation to linearize the dependency between the filtered components; and (c.) inspect the cross-dependency, including frequency-specific lead-lag structure through a vector autoregressive (VAR) model (\autoref{fig:graphical-summary}).

In studying frequency-driven cross-dependency between signals, one important operation is modulation, which is formally defined below in addition to a brief illustration in periodic signals.

\begin{definition}[Modulation]\label{def:modulation}Modulation is defined to be the product $y\left(n\right)$ of a signal $x_{m}\left(n\right)$ and another $x_{c}\left(n\right)$: %
\begin{equation}%
{y\left(n\right)=x_{1}\left(n\right)x_{c}\left(n\right)+\varepsilon\left(n\right)}%
\end{equation}%
 where $\varepsilon\left(n\right)$ is an i.i.d zero-mean white noise process.\end{definition}

\begin{remark}[Amplitude modulation]\label{rmk;dsb-am}When $x_{m}\left(n\right)$ and $x_{c}\left(n\right)$ are pure periodic waves with frequency $\omega_{m}$ and $\omega_{c}>\omega_{m}$, respectively, $x_{m}\left(n\right)=\cos\left(\omega_{m}n\right)$, $x_{c}\left(n\right)=\cos\left(\omega_{c}n\right)$. This modulation process is named as double-sideband suppress-carrier amplitude modulation (DSB-AM) in engineering \citep[ pp.~600]{SigSystems-Oppenheim-1997}, and $x_{c}$ is called the as carrier signal; $x_{m}$ is the modulator and $y$ is the modulated signal.\end{remark}

\subsection{Frequency filtering and nonlinear dependencies}

Frequency-selective filtering (or frequency filtering) is a process that has been well-studied in the signal processing literature for decomposing a signal into components with desired frequency properties \citep[ pp.~231-236]{SigSystems-Oppenheim-1997}. Each EEG recording at a channel is separated into several oscillations, and each one has spectra concentrated at pre-specified frequency bands. In this paper, we utilize a cascade Butterworth filter corresponding to the filtering method portrayed in Appendix \ref{sec:Appendix-Filters}.

Since each EEG can be characterized as a mixture of various oscillations, EEG dependence will be examined through the cross-dependence between those oscillatory components. These cross-frequency modulation effects, may be a real biological phenomenon supported by the findings of Sato et al. \citep{BiDirElec-Sato-2017}, Nozaradan et al. \citep{SteadystateEvokedPotentials-Nozaradan-2012}, Orekhova et al. \citep{FreqGammaOsc-Orekhova-2015}, and Albada-Robinson \citep{EEGSpectral-Albada-2013}. However, it is not straightforward to develop a causality framework based on a VAR modeling of the EEG oscillatory components. The two main issues are (a) nonlinear dependency across frequency bins and (b) low-frequency multicollinearity. We will develop a method that addresses these two crucial points. First, consider a bivariate setting below that demonstrates that a VAR model may fail to capture dependency -- even when it is present.

\begin{lemma}[]\label{thm:inter-freq}Consider 2-channel EEG driven by the following bivariate process%
\begin{equation}%
{
\begin{aligned}
x\left(n\right)
  & =\cos\left(2\pi\omega_{0}\left(n-1\right)\right)
+\varepsilon_{x}\left(n\right)\\
y\left(n\right) 
  & =\cos\left(2\pi\kappa\omega_{0}n\right)\cos\left(2\pi\omega_{0}n\right)
+\varepsilon_{y}\left(n\right)\end{aligned}
}%
\end{equation}%
where $\kappa>0$; $\varepsilon_{x}\left(n\right)$ and $\varepsilon_{y}\left(n\right)$ are zero-mean uncorrelated Gaussian white noise with variances $\sigma_{\varepsilon x}^{2}\left(n\right)$ and $\sigma_{\varepsilon y}^{2}\left(n\right)$, respectively. In this setting where there is clear dependence between $x$ and $y$, the VAR(2) model will show a negligible linear dependency $x\left(n\right)\mapsto y\left(n\right)$ for $\kappa\ge10$.\end{lemma}

\begin{proof}(Proof in Appendix \ref{sec:Appendix-VAR-modulation})\end{proof}

\autoref{thm:inter-freq} shows that binary relationships are highly dependent on the frequency of the pair of signals. It can be seen as a natural consequence of sinusoidal waves' orthogonality property in a multivariate regression context, and it can raise spurious dependency links under inter-frequency sinusoidal lagged interactions.

Therefore, any pair of signals may appear to be uncorrelated if they have a slightly different frequency. As a result, VAR models would fail to explain inter-frequency dependencies across channels. This premise can be extended to non-sinusoidal interactions. For instance, consider two time series $x_{A}\left(n\right)$, $x_{B}\left(n\right)$, modeled as VAR(2), which are contemporaneously dependent. That is, condition on $x_{B}\left(n\right)$, the time series $x_{A}\left(n\right)$ is uncorrelated with the past values $x_{B}\left(t\right)$ where $t = n-1, n-2, \ldots$. The VAR model would provide a false positive lagged correlation if the resonating frequencies of both time series are very low as it is shown in the following theorem.

\begin{lemma}[]\label{thm:correlation-frequency}Assume two zero-mean time series, $x_{A}\left(n\right)$ and $x_{B}\left(n\right)$, described through a second-order autoregressive model,%
\begin{equation}%
{
\begin{aligned}
x_{A}\left(n\right) & =\phi_{1A}x_{A}\left(n-1\right)+\phi_{2A}x_{A}\left(n-2\right)+\epsilon_{A}\left(n\right)\\
x_{B}\left(n\right) & =\phi_{1B}x_{B}\left(n-1\right)+\phi_{2B}x_{B}\left(n-2\right)+\epsilon_{B}\left(n\right)
\end{aligned}
}%
\end{equation}%
with correlated $\epsilon_{A}$ and $\epsilon_{B}$, resonating frequencies as $f_{A}^{*}$ and $f_{B}^{*}$, and recorded at a sampling frequency $f_{s}$. Both time series $x_{A}\left(n\right)$ and $x_{B}\left(n\right)$, present a cross-correlation at the first lag $\rho_{AB}\left(1\right)$ and $\rho_{BA}\left(1\right)$ given by%
\begin{equation}%
{
\begin{aligned}
\rho_{AB}\left(1\right) 
  & =2\rho_{AB}\left(0\right)
  \cos\left(2\pi\omega_{A}^{*}\right)
  \sqrt{-\phi_{2A}}
  \frac{1+\sqrt{\phi_{2A}\phi_{2B}}
          \frac{\cos\left(2\pi\omega_{B}^{*}\right)}
               {\cos\left(2\pi\omega_{A}^{*}\right)}
       }{1-\phi_{2A}\phi_{2B}}\\
\rho_{BA}\left(1\right)
  & =2\rho_{AB}\left(0\right)
 \cos\left(2\pi\omega_{B}^{*}\right)
 \sqrt{-\phi_{2B}}
 \frac{1+\sqrt{\phi_{2A}\phi_{2B}}
         \frac{\cos\left(2\pi\omega_{A}^{*}\right)}
         {\cos\left(2\pi\omega_{B}^{*}\right)}
      }{1-\phi_{2A}\phi_{2B}}
\end{aligned}
}%
\end{equation}%
where $\omega_{A}^{*}=\frac{f_{A}^{*}}{f_{s}}$ and $\omega_{B}^{*}=\frac{f_{B}^{*}}{f_{s}}$, $\rho_{AB}\left(0\right)$ is the correlation between $x_{A}$ and $x_{B}$.\end{lemma}

\begin{lemma}[Multicollinearity]\label{lem:multicollinearity}Both zero-mean time series, $x_{A}\left(n\right)$ and $x_{B}\left(n\right)$ , with identical resonating frequencies $f^{*}\le0.02f_{s}$ with frequency bandwidths $\tau\ge3$ have a cross-correlation at the first lag $\rho_{AB}\left(1\right)>0.991$.\end{lemma}

\begin{proof}(Proof in Appendix \ref{sec:Appendix-Multicollinearity}).\end{proof}

\autoref{lem:multicollinearity} also points out another issue in the interpretation of EEG signals. Common EEGs have sampling rates of 120Hz, 200Hz, 512Hz, up to 1000Hz depending on biomedical equipment settings. At a sampling frequency of 200Hz, and knowing that the maximum frequency of the delta rhythm is 4Hz ($\omega^{*}=0.02$), the autocorrelation for the first lag in these signals can be higher than 0.991 for this frequency band (under high signal-to-noise ratio conditions such $\tau\ge3$). Therefore, the estimates can be affected by both multicollinearity and identifiability issues. Autocorrelation can be overestimated, especially when the signals are obtained from higher sampling rates. Under the spectral causality framework, we now introduce a mapping procedure to prevent both conditions.

\subsection{Frequency mapping}\label{sec:Frequency-mapping}

Given that it is recognized that to perform a linear regression between frequency components, the decomposed time series to be compared should be in the same frequency range. However, the target EEG rhythms that are the building blocks of the observed EEG are obtained at different intervals by definition. Thus, Molaee et al. \citep{DeltaWavesDifferently-Molaee-Ardekani-2007} proposed to analyze (undirected) cross-frequency interactions translating all frequencies to an identical and bounded phase space. However, our proposed approach for comparing oscillations is to perform a frequency translation to map all signals to the same spectrum space. This will be explained below.

Consider the signal $x^{\left(i\right)}$ from channel $i$ which contains rhythms whose frequencies live exclusively in the interval $\left[f_{a},f_{b}\right]$ which is assumed to be sufficiently narrow by imposing that $f_{b}\le 3f_{a}$. Now, multiply the signal $x\left(t\right)$ with a cosine function of frequency $f_{a}$: %
\begin{equation}%
{x_{m}^{\left(i\right)}\left(n\right)=x^{\left(i\right)}\left(n\right)\cdot\cos\left(2\pi f_{a}n\right)}%
\end{equation}%
 Thus, $x_{m}^{\left(i\right)}\left(t\right)$ has its frequency components into two non-overlapping intervals $\left[0,f_{b}-f_{a}\right]$ and $\left[2f_{a},f_{b}+f_{a}\right]$. In order to keep only the components on the first segment, we can apply a 3-order cascade Butterworth filter (as defined in Appendix \ref{sec:Appendix-Filters}),\begin{equation}{
x_{f}^{\left(i\right)}\left(n\right)=LPF_{f_{a}}\left(x_{m}^{\left(i\right)}\left(n\right)\right)\left(n\right)
}%
\end{equation}%
Therefore, all channels $x_{f}^{\left(i\right)}\left(t\right)$ will have the same constrained frequency interval, allowing them to recognize linear relationships between them properly. Note that $x_{f}^{\left(i\right)}$ has a phase of 180 degrees in comparison with $x^{\left(i\right)}$ (as it can easily proved as a consequence of the modulation property of Fourier transforms). As \autoref{thm:correlation-frequency} support, OLS estimates could experience collinearity issues due to their low-frequency range with respect to the sampling frequency.

We suggested to include another step to map all the signals to an intermediate higher frequency $f_{i}$. This transformation will reduce the autocorrelations at lag 1, and therefore, lessen the multicollinearity issues. Additionally, this mapping process can be performed in a way that will compensate for the previous out-of-phase artifact. In this paper, we propose $f_{i}=0.2f_{s}$ as the intermediate target frequency, as \autoref{tab:cross-correlation} shows, the correlation is not higher than 0.810, even for pure sinusoidal waves.

This frequency translation is executed by creating a signal $x_{s}\left(n\right)$: %
\begin{equation}%
{x_{s}^{\left(i\right)}\left(n\right)=x_{f}^{\left(i\right)}\left(n\right)\cdot\cos\left(2\pi f_{i}n\right)}%
\end{equation}%
 As it was noticed before, this cosine-multiplication also introduces two harmonics into $\left[f_{i}-f_{b}+f_{a},f_{i}\right]$ and $\left[f_{i},f_{i}+f_{b}-f_{a}\right]$. We employ a band-pass filter on $\left[f_{i}-f_{b}+f_{a},f_{i}\right]$ to mitigate the out-of-phase issue generated in the previous step:\begin{equation}{
z_{\psi}^{\left(i\right)}\left(n\right)
= BPF_{f_{i}-f_{b}+f_{a}}
     ^{f_{i}}
  \left(
      x_{s}^{\left(i\right)}\left(n\right)
  \right)
  \left(n\right)
}\label{eq:mapped-freq}%
\end{equation}%
Finally, to assure that the same band of frequencies is being contrasted in the same frequency intervals, we split the spectrum into a set $\Psi$ of sub-bands with a constant frequency width of 4Hz (nomenclature of the divided subbands is shown in \autoref{tab:nomenclature-4Hz}). This frequency division is also motivated by the study developed by Albada et al. \citep{EEGSpectral-Albada-2013}, who split the beta band (4-8Hz) in two ranges, and found that each subband had different study behavior (the correlation of the spectrum peaks between each subband and the alpha band exhibit a significant difference).

\begin{table}\hypertarget{tab:nomenclature-4Hz}{%
\centering
\includegraphics[width=1\textwidth,height=\textheight]{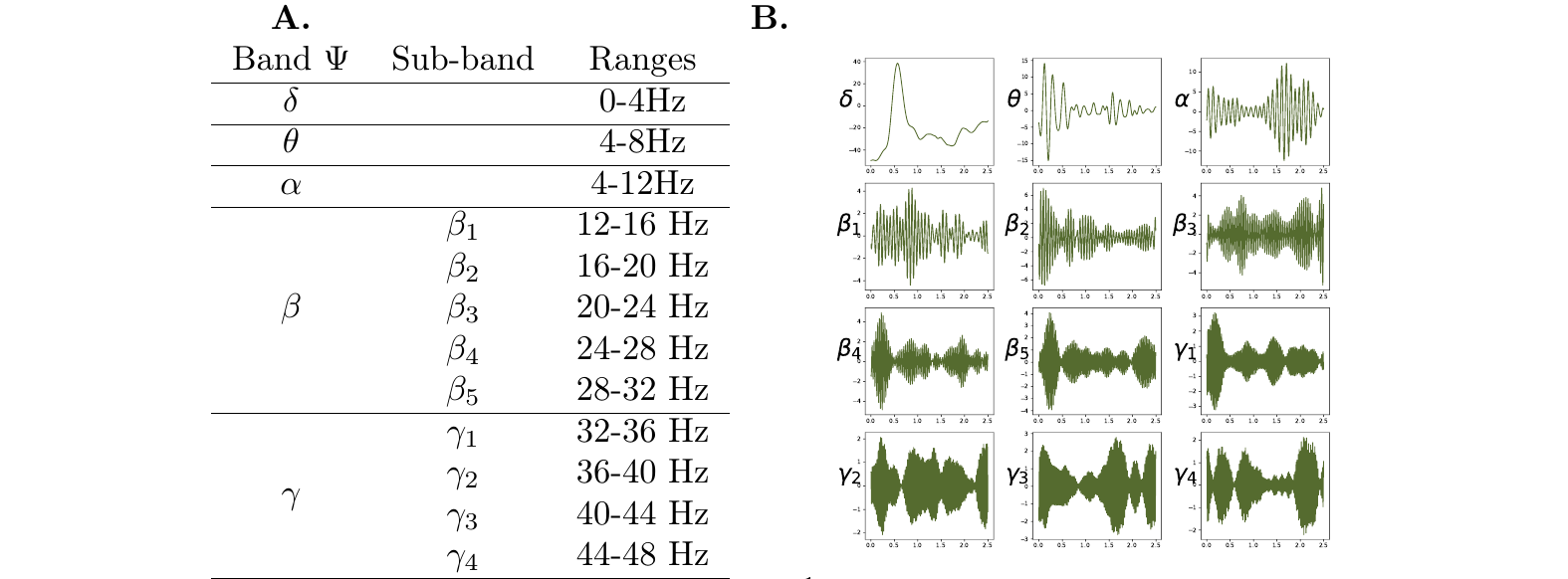}
\caption{4Hz-frequency divisions $\Psi$ for electroencephalographic signals. A) Nomenclature of the divided sub-bands. Note that $\beta_{5}$ contains frequencies at the boundaries of the beta and gamma rhythms. B) A sample of a frequency decomposition of a real EEG.}\label{tab:nomenclature-4Hz}
}
\end{table}

\subsection{Causality modeling}

With the preceding steps, the frequency relationships have been linearized, and we apply a VAR($p$) to model the dynamics of the $m$-channel multivariate observations $Z\left(n\right)=\left\{ \left\{ z_{\psi}^{\left(1\right)}\left(n\right)\right\} _{\psi\in\Psi},\left\{ z_{\psi}^{\left(2\right)}\left(n\right)\right\} _{\psi\in\Psi},\ldots,\left\{ z_{\psi}^{\left(m\right)}\left(n\right)\right\} _{\psi\in\Psi}\right\}$:%
\begin{equation}%
{
\begin{aligned}
Z\left(n\right) & =\sum_{\ell=1}^{p}\Phi_{\ell}Z\left(n-\ell\right)+\varepsilon\left(n\right)\\
\Phi_{\ell} & =\left(\begin{array}{cccccc}
\phi_{1,\alpha\leftarrow1,\delta}^{\left(\ell\right)} & \phi_{1,\alpha\leftarrow1,\theta}^{\left(\ell\right)} & \cdots & \phi_{1,\alpha\leftarrow m-1,\beta3}^{\left(\ell\right)} & \cdots & \phi_{1,\alpha\leftarrow m,\gamma4}^{\left(\ell\right)}\\
\phi_{2,\alpha\leftarrow1,\delta}^{\left(\ell\right)} & \phi_{2,\alpha\leftarrow1,\theta}^{\left(\ell\right)} & \cdots & \phi_{2,\alpha\leftarrow m-1,\beta3}^{\left(\ell\right)} & \cdots & \phi_{1,\alpha\leftarrow m,\gamma4}^{\left(\ell\right)}\\
\vdots & \vdots & \ddots & \vdots & \ddots & \vdots\\
\phi_{m,\alpha\leftarrow1,\delta}^{\left(\ell\right)} & \phi_{m,\alpha\leftarrow1,\theta}^{\left(\ell\right)} & \cdots & \phi_{m,\alpha\leftarrow m-1,\beta3}^{\left(\ell\right)} & \cdots & \phi_{1,\alpha\leftarrow m,\gamma4}^{\left(\ell\right)}
\end{array}\right)\nonumber \\
 & =\left(\begin{array}{c}
\phi_{i,\psi_{a}\leftarrow j,\psi_{b}}^{\left(\ell\right)}\end{array}\right)_{1\le i,j\le m;\psi_{a},\psi_{b}\in\Psi}\end{aligned}
}\label{eq:scau}%
\end{equation}%
Note that the system's dimensionality increases substantially with $144pm^{2}$ coefficients needed to be estimated. To provide an efficient regularization approach to estimate the model parameters, we rely on the method LASSLE proposed by Hu et al. \citep{HighMultichannel-Hu-2017}. The latter consists of executing the estimation in two phases: a) identify the relevant covariates using a LASSO regression, and b) use an ordinary least square to estimate the coefficients and their uncertainty.

\section{Spectral causality analysis of real EEG data}

\subsection{Data description}

\begin{figure}
\hypertarget{fig:wg-experiment}{%
\centering
\includegraphics[width=0.6\textwidth,height=\textheight]{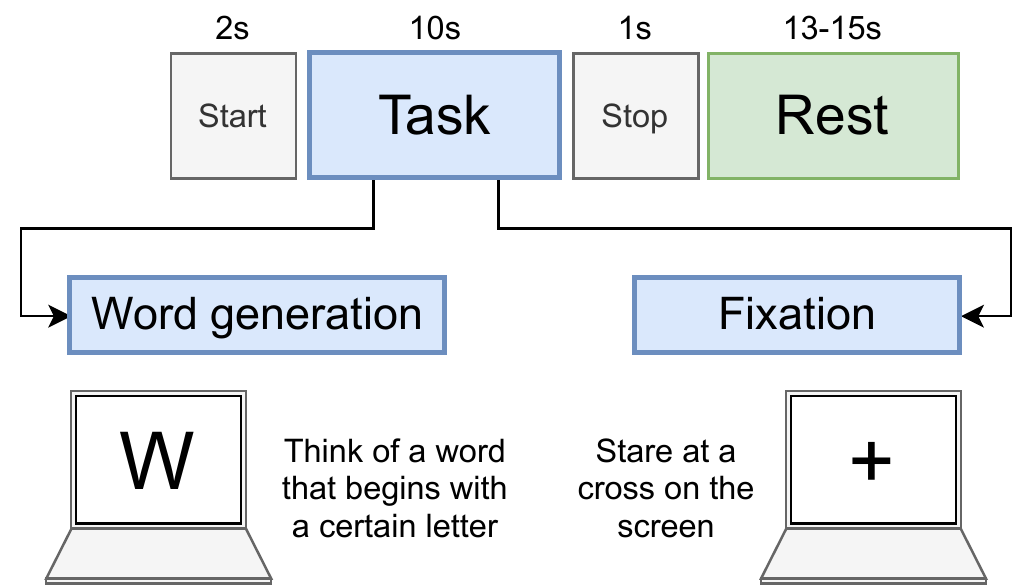}
\caption{Four steps of the experiment protocol for the word-generation task.}\label{fig:wg-experiment}
}
\end{figure}

Our spectral connectivity framework should properly describe brain dynamics when cross-frequency modulation effects are manifested in the signals. Given that certain categories of stimuli should trigger different networks of information flows in the brain, we can evaluate our framework by analyzing our method's potential to identify differences in the inferred networks.

In this paper, we adopt EEG recordings collected from Shin et al.~during a word generation experiment \citep{SimulAcquisitionEEG-Shin-2018} with 26 right-handed and healthy participants (9 males and 17 females) with an average of 26 years old. The dataset consists of 30 trials of word-generation (WG) tasks and 30 trials of fixation activities (FX) trials. In WG tasks, a single character is displayed to the participant while in FX tasks, a fixation cross is presented. Each trial consists of an instruction being portrayed for 2 seconds, followed by the execution of the task (WG or FX) for 10 seconds. Later, a stop signal accompanied by a short beep is raised for 1 second. Finally, there is a resting period from 13 to 15 seconds. \autoref{fig:wg-experiment} summarizes this protocol. The brain signals were collected using a BrainAmp EEG amplifier and an EASYCAP as a fabric cap, and with a sampling frequency of 200Hz. For further details about the dataset and the experimental protocol, we refer to \citep{SimulAcquisitionEEG-Shin-2018}.

\subsection{Data analysis method}

For the purposes of our study, we focus only on four channels: F1, F2, P7, and P8. The region covered by these channels is part of the predominant areas in which activation is reported during tasks related to lexical fluency \citep{CategoryLetterVerbal-Brickman-2005} or attention \citep{FrontalParietalAlpha-Misselhorn-2019} and is likely to contain relevant information for our analysis. To minimize the impact of eye and movement artifacts, the lagged effect of the vertical or horizontal electrooculogram signals in each EEG channel was removed \citep{RemovalOcularArtifact-Kenemans-1991}. Furthermore, the common signal across all channels (including those not included in our analysis) were subtracted.

Each channel was divided into 12 subbands and mapped into an intermediate frequency of $0.1f {s}=20$Hz, as described in Section~\ref{sec:Frequency-mapping}. Later, the signal was segmented in trials where the intervals related to the instruction message (first seconds) and the stop signal (one second after the task) were removed. For each step (\autoref{fig:wg-experiment}), the first 1000 points were used for estimating the system parameters.

In order to provide a background comparison, we estimated the connectivity networks obtained by using a linear vector autoregressive (VAR) model, as described in \autoref{eq:var-model}. Both models, SCAU and VAR, were fitted using the information from the previous 100ms, i.e., both models has a 20th order.

Within each model, the signal dynamics is determined by the set of coefficients $\{\Phi^{(VAR)}_\ell\}$ (\autoref{eq:var-model}) and $\{\Phi^{(SCAU)}_\ell\}$ (\autoref{eq:scau}) from the VAR and SCAU model, respectively. In order to provide an uniform connectivity measure, we rely on the global partial directed coherence (PDC) as a connectivity metric. PDC provides a measure of the information flow between two channels at a specific frequency $f$ using the coefficients of a VAR model:%
\begin{equation}%
{
\hat{\pi}_{i\mapsto j}^{(VAR)}\left(f\right)=\frac{A_{i\mapsto j}^{(VAR)}}{\sqrt{\sum_{j=1}^{m}\left|A_{i\mapsto j}^{(VAR)}\right|^{2}}}
}\label{eq:PDC-def}%
\end{equation}%
where $A_{i\mapsto j}^{(VAR)}=\delta_{ij}-\sum_{\ell=1}^{p}\hat{\phi}_{i,j}^{(VAR)}e^{-j2\pi f \ell}$. Recall that $\delta_{ij}$ is Kronecker delta, $m$ is the number of channels and $p$ is the order of the VAR model ($p=20$ in our analysis).

PDC is useful to quantify the frequency-dependent channel-to-channel effects, i.e., the total information flow from a certain frequency components of a source channel $i$ into a destination channel $j$. Therefore, the total PDC of the channel $i$ at a specific frequency interval $\psi=\left[f_0,f_1\right]$ ($\psi\in\Psi=\{\delta,\theta,\alpha,\beta_1,\ldots,\gamma_4\}$, \autoref{tab:nomenclature-4Hz} impacting a channel $j$ is defined by%
\begin{equation}%
{
\hat{I}_{i\mapsto j}^{(\psi,VAR)}
  =\int_{f_{0}}^{f_{1}}\left(\hat{\pi}_{i\mapsto j}^{(VAR)}\left(f\right)\right)^{2}\,df
}%
\end{equation}%
Note that PDC cannot identify specific frequency-to-frequency dependency effects. In contrast with VAR models, these type of dependencies are intrinsic embedded in SCAU models. Nevertheless, PDC can still provide an overall connectivity metric to quantify the information flow from the components with frequencies in $\psi_i$ in a channel $i$ to the components $psi_j$ in a channel $j$:%
\begin{equation}%
{
\hat{I}_{i\mapsto j}^{(\psi_{i},\psi_{j},SCAU)}
  =\int_{0}^{\nicefrac{1}{2}}
   \left(
     \hat{\pi}_{i,\psi_{i}\mapsto j,\psi_{j}}^{(SCAU)}
      \left(f^{*}\right)
   \right)^{2}\,df^{*}
}%
\end{equation}%
In this formulation, the PDC estimator, $\hat\pi$, has been slightly adapted to the SCAU notation (\autoref{eq:scau}):%
\begin{equation}%
{
\hat{\pi}_{,i,\psi_{i}\mapsto j,\psi_{j}}^{(SCAU)}\left(f^{*}\right)
  =\frac{A_{i,\psi_{i}\mapsto j,\psi_{j}}^{(SCAU)}}{\sqrt{\sum_{j=1}^{m}\sum_{\psi_{j}\in\Psi}\left|A_{,\psi_{i}\mapsto j,\psi_{j}}^{(SCAU)}\right|^{2}}}
}%
\end{equation}%
where $A_{i,\psi_{i}\mapsto j,\psi_{j}}^{(SCAU)}=\delta_{ij}-\sum_{\ell=1}^{p}\hat{\phi}_{,i,\psi_{i}\mapsto j,\psi_{j}}e^{-j2\pi\omega f^{*}}$ and $f^{*}$ is a normalized mapped frequency $f^{*}\in[0,\nicefrac{1}{2}]$.

\subsection{Comparison of connectivity changes}

Both connectivity measures, $\hat{I}_{i\mapsto j}^{(\psi_{i},\psi_{j},SCAU)}$ and $\hat{I}_{i\mapsto j}^{(\psi,VAR)}$ serves a foundation for comparing any variation of brain connectivity networks from both modeling perspectives. However, we should consider time-varying effects that can modify the background connectivity networks. Therefore, we focus only into the relative connectivity $c$ that is defined as the difference in the brain network during a particular task that occur with respect to its following resting period, i.e.~$c^{(\cdot)}_{(\cdot)} = I^{(\cdot)}_{(\cdot)}\vert T_{WG} - I^{(\cdot)}_{(\cdot)}\vert T_{REST-WG}$ for the WG task.

As it was expressed before, we are interested in the ability of SCAU to find and quantify differences in the connectivity networks between WG and FX tasks across subjects and trials. Therefore, we define the contrast metric $d$ that measure the absolute difference, multiplied by 100, between the connectivity from the frequency components in the interval $\psi_i$ within a channel $i$ towards the components in $\psi_j$ in a channel $j$:%
\begin{equation}%
{
d_{i\leftarrow j}^{(\psi_{i},\psi_{j})} 
  = 100 \left|
    c_{i\leftarrow j}^{(\psi_{i},\psi_{j})}
   -c_{i\leftarrow j}^{(\psi_{i},\psi_{j})}
  \right|
}%
\end{equation}%

\begin{figure}
\hypertarget{fig:C2C}{%
\centering
\includegraphics[width=0.6\textwidth,height=\textheight]{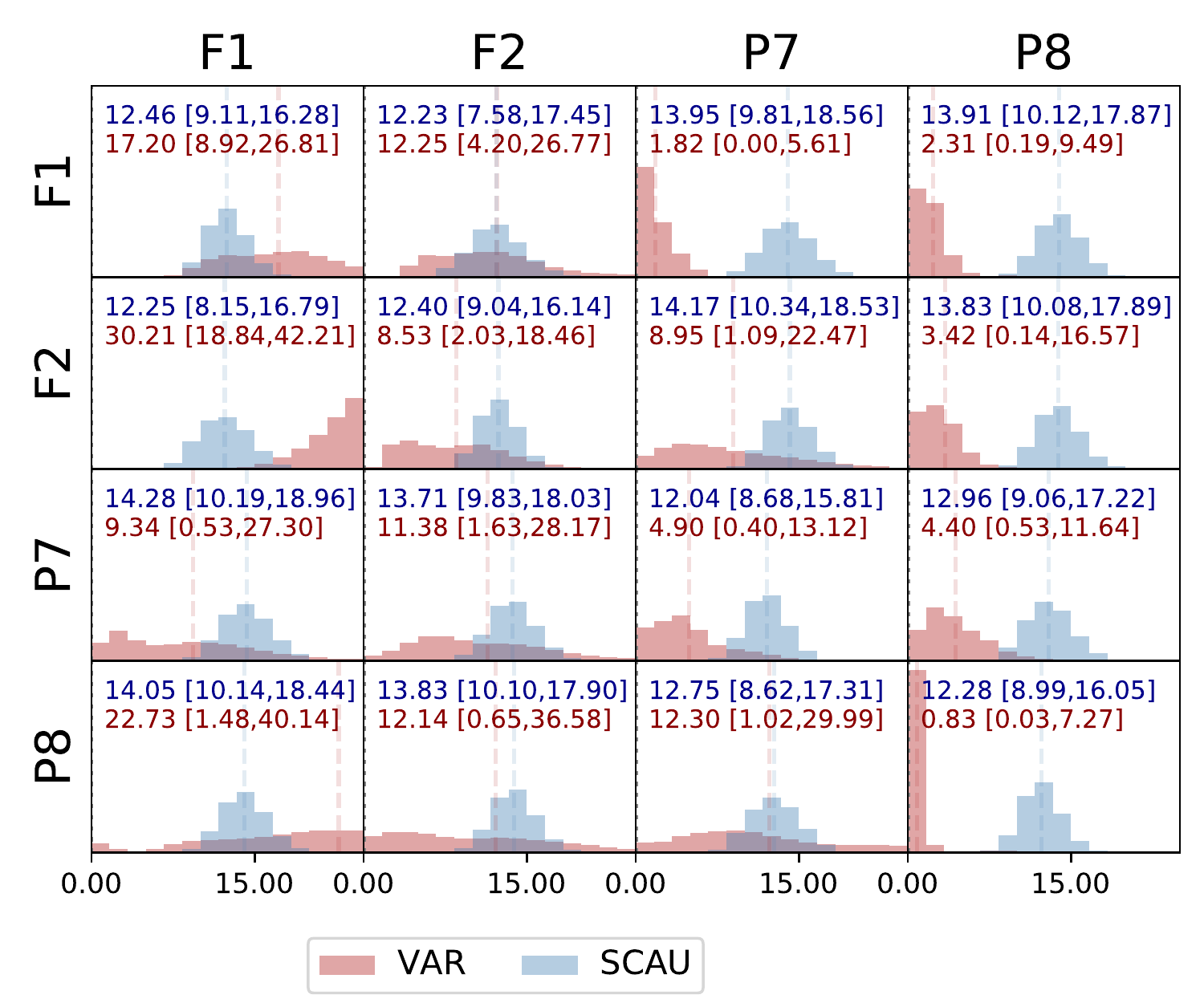}
\caption{Channel-to-channel interactions: Bootstrap distribution of the mean contrast between WG and FX stimuli by using the VAR and SCAU model. The mean value and its 95\% confidence interval are shown in the labels for each possible combination.}\label{fig:C2C}
}
\end{figure}

\begin{figure}
\hypertarget{fig:F2C}{%
\centering
\includegraphics[width=1\textwidth,height=\textheight]{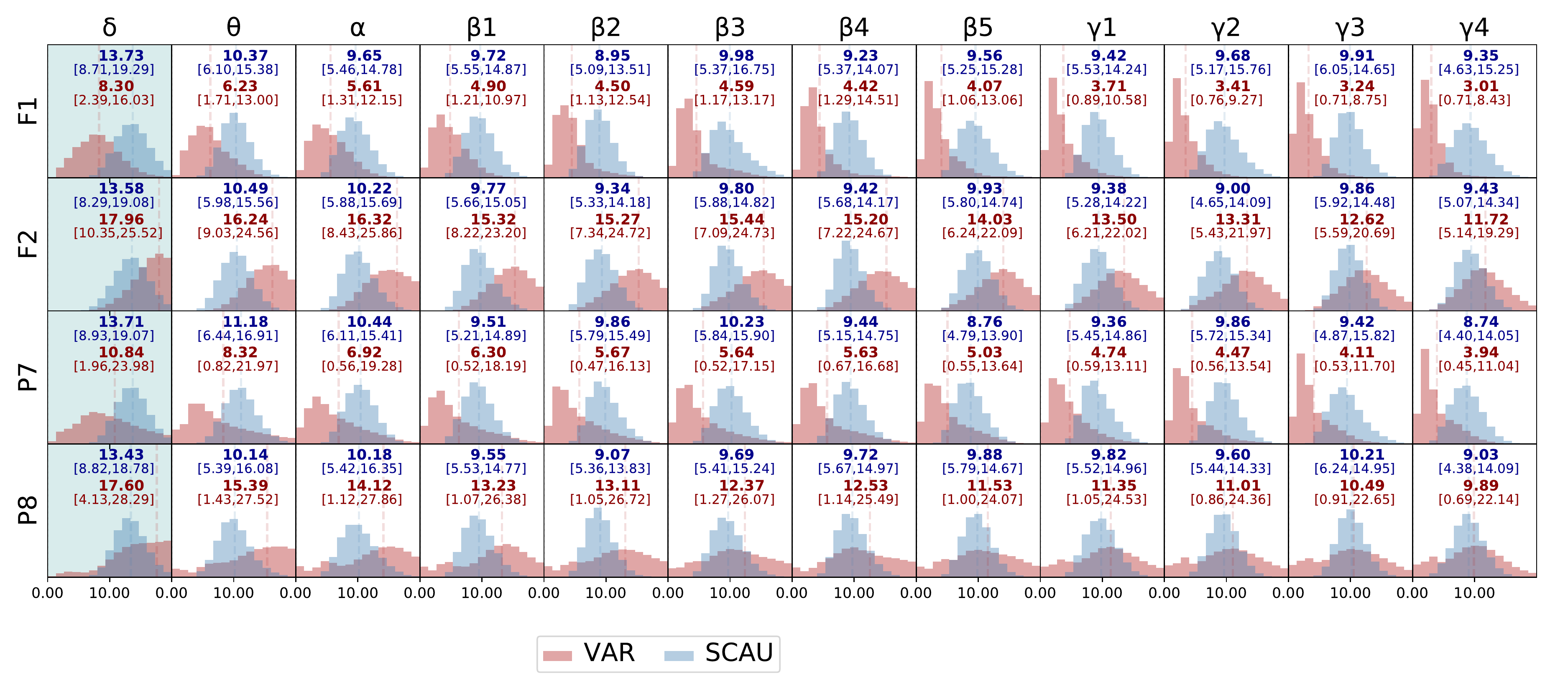}
\caption{Contrast in information flows originated in a frequency interval towards a channel (frequency-to-channel). Each cell shows the cross-subject bootstrap distributions of the mean contrast of relative connectivity networks estimated by VAR (red) and SCAU (blue) models. Note that VAR-estimated contrasts show either a high variance or mean values close to zero. On the contrary, SCAU contrast shows a lower variance distribution with higher mean contrasts on the interactions where signals are originated in the delta band.}\label{fig:F2C}
}
\end{figure}

\begin{figure}
\hypertarget{fig:C2F}{%
\centering
\includegraphics[width=1\textwidth,height=\textheight]{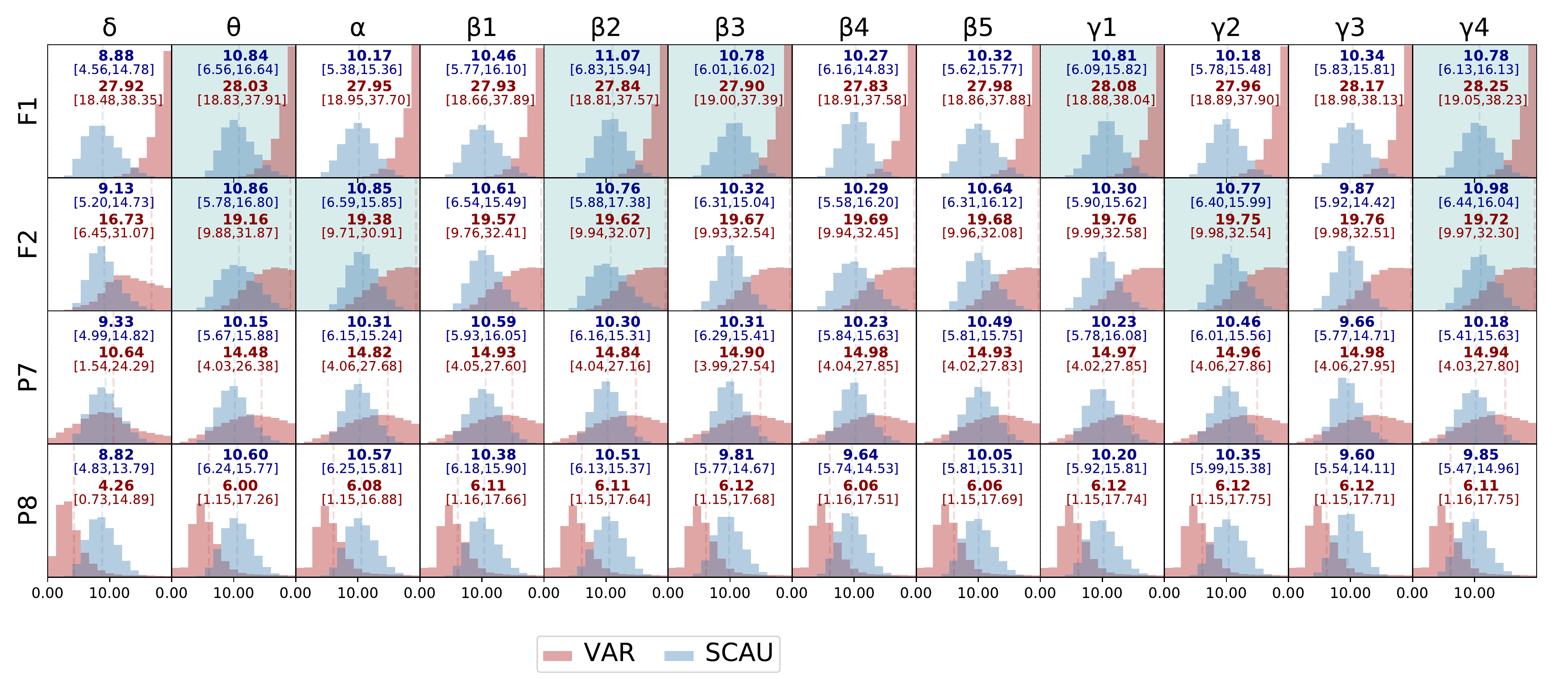}
\caption{Contrast in information flows originated in a channel towards a frequency interval (channel-to-frequency). Each cell shows the cross-subject bootstrap distributions of the mean contrast of relative connectivity networks estimated by VAR (red) and SCAU (blue) models. VAR models estimates information flows at the frontal channels with higher magnitudes in comparison with the SCAU model. Note that both models estimates similar information flows caused by channel P8.}\label{fig:C2F}
}
\end{figure}

\begin{figure}
\hypertarget{fig:summary-net}{%
\centering
\includegraphics[width=0.7\textwidth,height=\textheight]{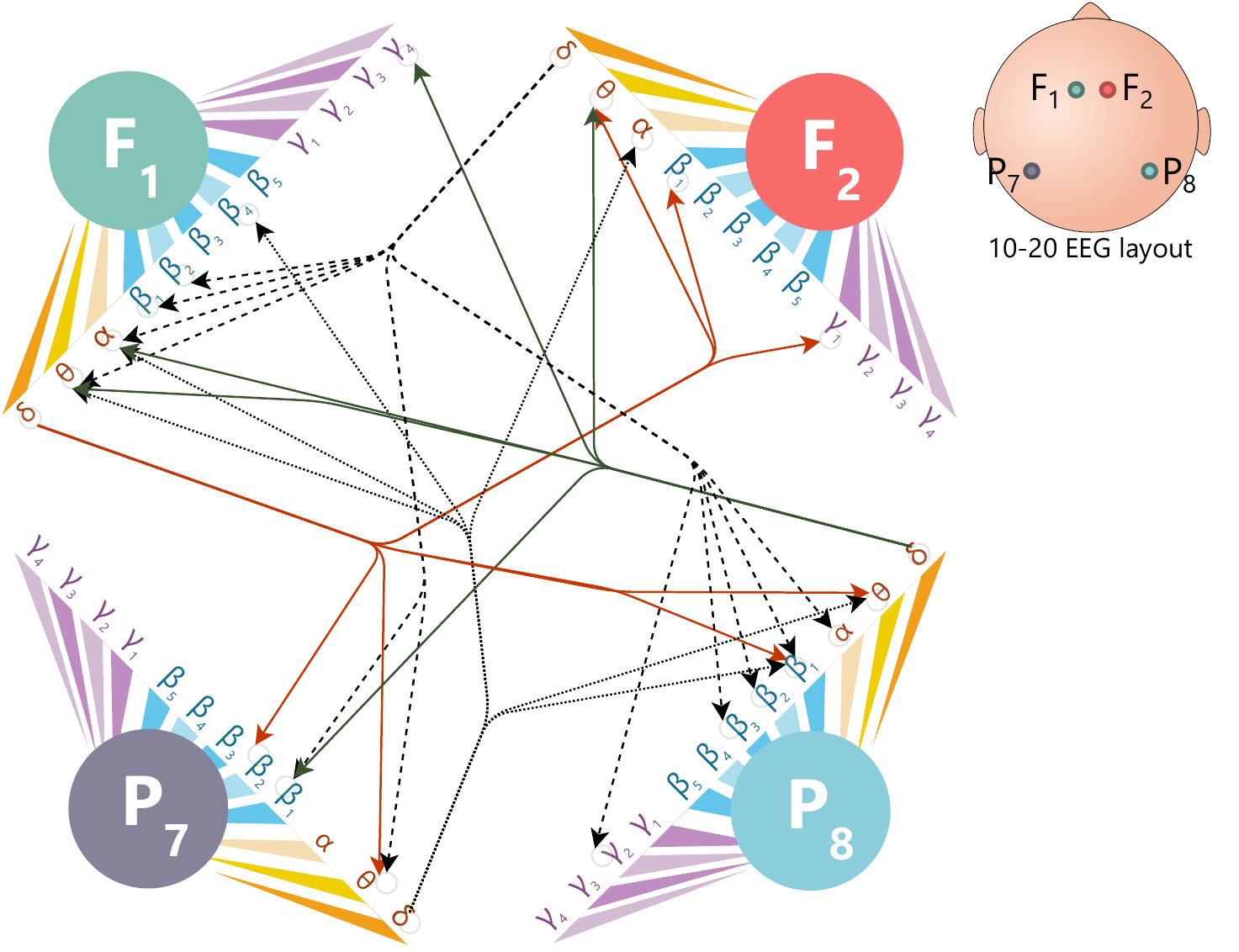}
\caption{Summary cross-subject contrast information flows across channels and frequency-channels. Connections with mean contrast higher than 80\% of the maximum contrasts are denoted. Highlight the source of links in the delta band and high level of connectivity towards the alpha and beta frequency components.}\label{fig:summary-net}
}
\end{figure}

\section{Results and discussion}

\textbf{Multilevel SCAU interpretability.} One of the advantages of SCAU over other frequency-causality models is its ability to explain dependence from several perspectives by modeling and quantifying spectro-spatial relationships. It is, therefore, possible to recognize at least five levels of interpretation that SCAU can provide to any estimated network: (A) overall channel dependence (channel-to-channel interactions); (B) overall frequency band dependence (frequency-to-frequency modulation); (C) spectral impact towards channels (frequency-to-channel dependence) and (D) spectral influence of each channel (channel-to-frequency dependence), and (E) Spatio-spectral dependency (full channel-frequency interactions)

As it was stated before, our analysis was conducted in EEG data from 26 subjects, while they were performing two tasks: word generation (WG) and visual fixation (FX) that were repeated during 30 trials. In view of the large amount of data and network parameters, the analysis and interpretation focuses on three types of interactions: a) channel-to-channel (C2C), b) frequency-to-channel (F2C), and c) full frequency-channel (FC2FC). Therefore, the spectro-spatial network of information flows provided by SCAU can be contrasted with VAR-PDC metrics, and previous studies can be examined for biological interpretation.

The bootstrap distribution for the F2C and C2C interactions are shown in \autoref{fig:F2C} and \autoref{fig:C2C}, while a summary diagram for the third type of analysis (FC2FC) case is shown in \autoref{fig:summary-net}.

\textbf{Channel-to-channel dependency flows.} Let us first consider all interactions that are originated from a channel that impacts another, i.e., C2C dependency flows. Both models, VAR and SCAU are capable of representing such types of dependence.

Our results shows that, using a VAR model, the changes in the relative connectivity between lexical (WG) and fixation (FX) tasks fall into two recognizable categories: (a.) C2C interactions are not significantly different given the 95\% confidence interval contains the null case (e.g., $P8 \mapsto F2$); or (b.) a change is observed but it has a large variance (e.g., $F1\mapsto F2$, $F1 \mapsto P8$, $F2 \mapsto F1$). These categories are easily recognizable in \autoref{fig:C2C}.

On the other side, we can observe that SCAU allows us to identify a significant difference with lower variance across all possible channel interactions. Furthermore, the strongest connections seem to be cross-hemispheric: $F1 \mapsto P7$ (14.27), $P7 \mapsto F2$ (14.17) and $F1 \mapsto P8$ (14.05).


\textbf{Frequency-to-channel and channel-to-frequency dependency flows.} We rearranged the links in order to reveal the impact of the frequency components on the channels (frequency-to-channel, F2C, modulation), and also, the channel effect towards the frequencies (channel-to-frequency, C2F, modulation). F2C and C2F analyze offer a complementary perspective to the classical analysis of inter-channel interactions.

As in the previous case, it is clear that the variance of the contrast estimator is higher when using a VAR model for modeling the signal dynamics. For instance, consider the notable mean effect of $\beta_2\mapsto{}P_8$: 13.11, with a confidence interval ranging from 0.85 to 29.67. High volatility, in consequence, reduces the ability to differentiate WG and FX across trials and subjects. In contrast, the SCAU model shows lower contrast but with narrow confidence intervals that are further from the null alternative. In the same $\beta_2\mapsto{}P_8$ link SCAU estimates a mean difference of 9.07 with a confidence interval from 3.62 to 16.64.

Furthermore, the strongest F2C interactions were originated at the $\delta$ band (with mean values higher than 6.0), while the most relevant C2F interactions involved the $\theta$, $\alpha$, $\beta_{2-3}$ and $\gamma_{1,2,4}$ bands as signal receivers with mean values higher than 10.75. The involvement of $\delta$, $\theta$, and $\gamma$ are related to visual and attention activities: highly-synchronized interactions in higher regions of the $\gamma$ band (36-56Hz) involving fronto-parietal links have been described to be related to visual searching tasks \citep{FrontalParietalSynchrony-Phillips-2010}. Moreover, dependence flows in the $\theta$ band have been described as patterns related to attention or focus in visual and auditory tasks \citep{FrontalParietalAlpha-Misselhorn-2019}. Harper et al. \citep{ThetaDeltabandEEG-Harper-2017} also described that dependence links in the delta and theta waves in the frontoparietal region could be related to attention and stimulus detection \citep{ThetaDeltabandEEG-Harper-2017}. We should emphasize that these studies did not consider the directionality of the spectro-spatial dependency, i.e., the model does not differentiate if the frequency component leads the flow towards the channel or vice versa. As a particular case of analysis, the estimated mean SCAU contrast denotes that the links with the highest connectivity are originated in the delta band (\autoref{fig:F2C}) while they are received in theta and high gamma ($\gamma_4$, 24-28Hz) components (\autoref{fig:C2F}). However, further experiments should be developed in order to confirm this dependence directionality.

\textbf{Full frequency-channel dependency map.} As it was previously mentioned, in contrast with the VAR model, SCAU can capture a full description of the frequency-channel flows of information. In our dataset, this implies 2304 estimators as a result of the combination of source-detector flows generated by 12 bands and four channels. In order to summary this large number of estimators, we only focus on the strongest information flows. For our purpose, those links are defined as the flows with mean contrast is higher or equal to 80\% of the maximum contrast. \autoref{fig:summary-net} shows a network map of these links.

We emphasize that the summary network highlighted the role of the delta band as a predominant signal modulator. We should recall that delta waves are significantly associated with mental calculation and concentration, while their decrease in power could also be associated with ``attention to the external world'' as a stimulus trigger \citep{FSignificanceDelta-Harmony-2013}.

Moreover, a distinctive delta-theta modulation was detected across the four channels. There is no definitive cause for this type of modulation phenomenon to the best of our knowledge, but it is known that changes in the common (phase-amplitude) delta-theta patterns can be caused due to anesthetic effects \citep{DeltaWavesDifferently-Molaee-Ardekani-2007}.

\section{Conclusions}

Modulation seems to be a natural phenomenon that occurs under different contexts in EEG. Signal modulation in the beta (12-30Hz) and gamma band (30-50Hz) has been reported as a result of visual stimuli \citep{FreqGammaOsc-Orekhova-2015, BetaOscOsc-Piantoni-2010}, visual information processing \citep{BiDirElec-Sato-2017}, as well as a consequence of the effect of external signal sources such as speed-variable inputs \citep{FreqGammaOsc-Orekhova-2015} or frequency-variable music \citep{SteadystateEvokedPotentials-Nozaradan-2012}. In this paper, we introduce a time series model to capture time-causality (or Granger-causality) EEG dynamics at a channel-frequency (or spectro-spatial) level. Our proposed spectral causality model (SCAU) can capture information flows that can take place among the recorded channels but also covering spectral connections that could have been masked due to frequency modulation effects.

Our method is performed in three fundamental phases: (A.) Frequency decomposition, where each recorded channel is split into non-overlapping frequency bands with some intrinsic biological explanation. (B.) Frequency mapping where all signals are translated into a common mid-frequency space to linearize the cross-frequency dependencies and mitigate cross-frequency interaction issues. (C.) Dependency modeling using a vector autoregressive model (VAR) on the decomposed and mapped time series using a specific regularization method (LASSLE).

We applied the SCAU into an EEG dataset from twenty-six participants performing two visual/lexical tasks: word generation and visual fixation during 30 trials. The estimated SCAU connectivity networks were compared with networks estimated by a Vector Autoregressive model (VAR). We analyzed the bootstrap distribution of the mean difference (or contrast) between both networks and proposed a multilevel spectro-spatial interpretation through channel-to-channel, frequency-to-channel, channel-to-frequency, and full channel-frequency interactions. A non-null consistent contrast was detected across trials and subjects, allowing us to denote the better capability of the SCAU model to differentiate the connectivity networks for both stimuli (in comparison with the VAR-estimated networks). Moreover, the estimated connectivity was consistent with the expected response discussed in the literature, i.e., stronger interactions in theta and alpha bands were observed during word generation and visual fixation, respectively. In addition, unreported effects of cross-frequency modulation have also been found using SCAU.

The enhanced predictive capacity and high interpretability of results allowed us to propose our SCAU model as a potential alternative method to VAR models for characterizing time-causal patterns embedded in multivariate time series. We also highlight that SCAU can be applied to any other type of signal where frequency components have an interpretable and biological function and are presumed to interact among them.

\begin{appendices}
\clearpage
\section{Butterworth filtering design}\label{sec:Appendix-Filters}

\subsection{Low-pass filter}

Let $f_{c}$, $f_{s}$, and $N$ be the cut-off frequency, sampling frequency, and the filter order, respectively. Analog Butterworth filters do not have zero components, but only poles defined by \citep[pp.~629-632]{AppliedDigitalSig-DimitrisGManolakis-2011}:%
\begin{equation}%
{
p=\left\{ -e^{j\pi\frac{m}{N}}\vert m
  =-\left\lfloor \frac{N-1}{2}\right\rfloor \ldots\left\lfloor \frac{N}{2}\right\rfloor \right\}
}%
\end{equation}%
In order to produce a digital filter from the analog design, we define the normalized warped frequency%
\begin{equation}%
{
\omega_{o}=4\tan\left(\pi\frac{f_{c}}{f_{s}}\right)
}%
\end{equation}%
with the normalized gain%
\begin{equation}%
{
k_{o}=\omega_{o}^{\left|p\right|}
}\label{eq:gain-LP}%
\end{equation}%
where $\left|p\right|$ is the length of the vector $p$.

Therefore, the adjusted poles of the digital filter are given by%
\begin{equation}%
{
p_{o}=w_{o}p
}\label{eq:poles-LP}%
\end{equation}%
Finally, the filter would be characterized by the transfer function%
\begin{equation}%
{
H\left(z^{-1}\right)
  =\frac{k_{o}}
        {\prod_{i=1}^{\left|p\right|}
           \left(
             z^{-1}-w_{o}p_{i}
           \right)}
}%
\end{equation}%

\subsection{Band-pass filter}

Let $f_{s}$, $N$, $f_{c1}$,$f_{c2}$ be the sampling frequency, filter order, and the lower and upper cut-off frequencies, respectively. The design of the band-pass filter will be obtained as a variation of the low-pass filter design. Assume the same vector of analog poles $p$, and, consider a normalized warped equivalent for each cut-off frequency:%
\begin{equation}%
{
\begin{aligned}
  \omega_{o1} 
    & =4\tan\left(\pi\frac{f_{c1}}{f_{s}}\right)\\
  \omega_{o2}
    & =4\tan\left(\pi\frac{f_{c2}}{f_{s}}\right)
\end{aligned}
}%
\end{equation}%
Later, define a central warped frequency%
\begin{equation}%
{
\omega_{o}=\sqrt{\omega_{o1}\omega_{o2}}
}%
\end{equation}%
and a spectral bandwidth%
\begin{equation}%
{
b_{\omega}=\omega_{o2}-\omega_{o1}
}%
\end{equation}%
The normalized gain would be defined in a similar structure as the low-pass design%
\begin{equation}%
{
k_{o}^{BP}=\omega_{o}^{\left|p\right|}
}\label{eq:gain-BP}%
\end{equation}%
But, the vector of poles will be expressed as the concatenation of two adjusted pole vectors:%
\begin{equation}%
{
p_{o}
  =\left\{
      \frac{b_{\omega}}{2}p
     +\frac{b_{\omega}}{2}\sqrt{p^{2}-\omega_{o}^{2}}
     ,
      \frac{b_{\omega}}{2}p
     -\frac{b_{\omega}}{2}\sqrt{p^{2}-\omega_{o}^{2}}
   \right\}
}\label{eq:poles-BP}%
\end{equation}%
Under this notation, the filter is also described by an equivalent transfer function%
\begin{equation}%
{
H\left(z^{-1}\right)
  =\frac{k_{o}}
        {\prod_{i=1}^{\left|p\right|}
          \left(z^{-1}-w_{o}p_{i}\right)
        }
}%
\end{equation}%

\subsection{Cascade implementation}

For numerical stability, we rely on the transposed direct form II implementation of digital filters in this paper. First, we construct a set of coefficients $a=\left\{ a_{1},a_{2},\ldots\right\}$ obtained as the polynomial coefficients of the transfer function's denominator \citep[pp.~496-500]{AppliedDigitalSig-DimitrisGManolakis-2011}:%
\begin{equation}%
{
\prod_{i=1}^{\left|p\right|}\left(z^{-1}-w_{o}p_{i}\right)
  = \sum_{i=1}^{N}a_{i}z^{-i}
}\label{eq:polynomial-coeff}%
\end{equation}%
Therefore, the standard difference equation for the filter can be written as%
\begin{equation}%
{
\begin{aligned}
  \nu\left(t\right)
    & = y\left(t\right)
        -\sum_{i=1}^{N}a_{i}v\left(t-\ell\right)\\
  \tilde{y}\left(t\right)
    & =k_{o}\nu\left(t\right)
\end{aligned}
}%
\end{equation}%
where $y\left(t\right)$ is the input signal, $\nu\left(t\right)$ is the latent component, $\tilde{y}\left(t\right)$ is the output filtered signal.

However, due to the absence of zeros in Butterworth filters, we can rewrite the difference equation as%
\begin{equation}%
{
\begin{aligned}
  \tilde{y}\left(t\right)
    & = k_{o}y\left(t\right)
      -\sum_{i=1}^{N}a_{i}\tilde{y}
       \left(t-\ell\right)
\end{aligned}
}%
\end{equation}%
We should emphasize that higher-order filters would offer better attenuation in the rejection band at the cost of likely instability if the poles are closer to the unit circle. Thus, in this paper, we used a 3-level cascade structure for low order filters:

\begin{equation}%
{
\begin{aligned}
  \tilde{y}\left(t\right)
  & =-\sum_{\ell=1}^{N}a_{\ell}\tilde{y}
      \left(t-\ell\right)+k_{o}\tilde{y}_{1}\left(t\right)
      \\
  \tilde{y}_{1}\left(t\right)
  & =-\sum_{\ell=1}^{N}a_{\ell}\tilde{y}_{1}
      \left(t-\ell\right)+k_{o}\tilde{y}_{2}\left(t\right)\nonumber \\
  \tilde{y}_{2}\left(t\right) 
  & =-\sum_{\ell=1}^{N}a_{\ell}\tilde{y}_{2}
      \left(t-\ell\right)+k_{o}y\left(t\right)\nonumber
\end{aligned}
}\label{eq:cascade-filter}%
\end{equation}%
where $\tilde{y}_{1}\left(t\right)$ and $\tilde{y}_{2}\left(t\right)$ are the intermediate filtering signals.

To simplify the notation of this process, we define the function%
\begin{equation}%
{
LPF_{f_{c}}\left(y\right)\left(t\right)
=\tilde{y}\left(t\right)
}\label{eq:lpf-filter}%
\end{equation}%
where $\tilde{y}\left(t\right)$ is calculated in \autoref{eq:cascade-filter} with the coefficients $a_{1},a_{2},\ldots a_{N}$ and $k_{o}$ defined by \Autoref{eq:polynomial-coeff,eq:gain-LP,eq:poles-LP}. To simplify the notation, we define the ``band-pass operator'' as%
\begin{equation}%
{
BPF_{f_{c1}}^{f_{c2}}\left(y\right)\left(t\right)
=\tilde{y}\left(t\right)
}\label{eq:bpf-filter}%
\end{equation}%
with $\tilde{y}\left(t\right)$ as \autoref{eq:cascade-filter} with the coefficients and gain established by \Autoref{eq:polynomial-coeff,eq:gain-BP,eq:poles-BP}.

\section{VAR and modulation}\label{sec:Appendix-VAR-modulation}

\begin{lemma*}[Nonlinear dependence]\label{thm:inter-freq0}Consider 2-channel EEG driven by the following bivariate process%
\begin{equation}%
{
\begin{aligned}
x\left(n\right)
  & =\cos\left(2\pi\omega_{0}\left(n-1\right)\right)
+\varepsilon_{x}\left(n\right)\\
y\left(n\right) 
  & =\cos\left(2\pi\kappa\omega_{0}n\right)\cos\left(2\pi\omega_{0}n\right)
+\varepsilon_{y}\left(n\right)\end{aligned}
}%
\end{equation}%
where $\kappa>0$; $\varepsilon_{x}\left(n\right)$ and $\varepsilon_{y}\left(n\right)$ are zero-mean uncorrelated Gaussian white noise with variances $\sigma_{\varepsilon x}^{2}\left(n\right)$ and $\sigma_{\varepsilon y}^{2}\left(n\right)$, respectively. In this setting where there is clear dependence between $x$ and $y$, the VAR(2) model will show a negligible linear dependency $x\left(n\right)\mapsto y\left(n\right)$ for $\kappa\ge10$.\end{lemma*}

\begin{proof}Assume a VAR(2) model of the joint observation $\left(x\left(n\right),y\left(n\right)\right)$:%
\begin{equation}%
{
\begin{aligned}
\left(\begin{array}{c}
x\left(n\right)\\
y\left(n\right)
\end{array}\right) & =\left(\begin{array}{cc}
x\left(n-1\right) & y\left(n-1\right)\\
x\left(n-1\right) & y\left(n-1\right)
\end{array}\right)\left(\begin{array}{cc}
\phi_{xx}^{\left(1\right)} & \phi_{xy}^{\left(1\right)}\\
\phi_{yx}^{\left(1\right)} & \phi_{yy}^{\left(1\right)}
\end{array}\right)\nonumber \\
 & +\left(\begin{array}{cc}
x\left(n-2\right) & y\left(n-2\right)\\
x\left(n-2\right) & y\left(n-2\right)
\end{array}\right)\left(\begin{array}{cc}
\phi_{xx}^{\left(2\right)} & \phi_{xy}^{\left(2\right)}\\
\phi_{yx}^{\left(2\right)} & \phi_{yy}^{\left(2\right)}
\end{array}\right)+\left(\begin{array}{c}
\varepsilon_{x}\left(n\right)\\
\varepsilon_{y}\left(n\right)
\end{array}\right)
\end{aligned}
}%
\end{equation}%
Now, let us estimate the set of coefficients $\phi_{xx}^{\left(1\right)},\phi_{xy}^{\left(1\right)},\phi_{xx}^{\left(2\right)},\phi_{xy}^{\left(2\right)}$ using least squares:%
\begin{equation}%
{
y\left(n\right)=\left(\begin{array}{cccc}
x\left(n-1\right) & x\left(n-2\right) & y\left(n-1\right) & y\left(n-2\right)\end{array}\right)\left(\begin{array}{c}
\phi_{xx}^{\left(1\right)}\\
\phi_{xx}^{\left(2\right)}\\
\phi_{xy}^{\left(1\right)}\\
\phi_{xy}^{\left(2\right)}
\end{array}\right)+\varepsilon_{y}\left(n\right)
}%
\end{equation}%
So the estimates are described by%
\begin{equation}%
{
\begin{aligned}
\left(\begin{array}{c}
\hat{\phi}_{xx}^{\left(1\right)}\\
\hat{\phi}_{xx}^{\left(2\right)}\\
\hat{\phi}_{xy}^{\left(1\right)}\\
\hat{\phi}_{xy}^{\left(2\right)}
\end{array}\right) & =\left(\begin{array}{cccc}
s_{n-1,n}^{(x,x)} & s_{n-1,n-2}^{(x,x)} & s_{n-1,n-1}^{(y,x)} & s_{n-2,n-1}^{(y,x)}\\
s_{n-1,n-2}^{(x,x)} & s_{n-2,n-2}^{(x,x)} & s_{n-1,n-2}^{(y,x)} & s_{n-2,n-2}^{(y,x)}\\
s_{n-1,n-1}^{(x,y)} & s_{n-2,n-1}^{(x,y)} & s_{n-1,n-1}^{(y,y)} & s_{n-2,n-1}^{(y,y)}\\
s_{n-1,n-2}^{(x,y)} & s_{n-2,n-2}^{(x,y)} & s_{n-1,n-2}^{(y,y)} & s_{n-2,n-2}^{(y,y)}
\end{array}\right)^{-1}\left(\begin{array}{c}
s_{n-1,n}^{(x,y)}\\
s_{n-2,n}^{(x,y)}\\
s_{n-1,n}^{(y,y)}\\
s_{n-2,n}^{(y,y)}
\end{array}\right)
\end{aligned}
}%
\end{equation}%
where $s_{n-2,n-1}^{(u,v)}$ is a condensed notation for $\text{cov}\left(u\left(i\right),v\left(j\right)\right)$.

From the generating model, we can calculate the covariance assuming an integration period of a cycle and a high sampling frequency in comparison with the resonating frequency $\kappa\omega_{0}$:%
\begin{equation}%
{
\begin{aligned}
x\left(n\right) & =\cos\left(2\pi\omega_{0}\left(n-1\right)\right)+\varepsilon_{x}\left(n\right)\\
y\left(n\right) & =\cos\left(2\pi\kappa\omega_{0}n\right)\cos\left(2\pi\omega_{0}n\right)+\varepsilon_{y}\left(n\right)
\end{aligned}
}%
\end{equation}%
\begin{equation}%
{
\begin{aligned}
\text{cov}\left(x\left(n-1\right),y\left(n-1\right)\right) & =\frac{1}{2\pi}\int_{0}^{2\pi}\left(\cos\left(2\pi\kappa\omega_{0}\nu\right)\cos^{2}\left(2\pi\omega_{0}\nu\right)\right)\,d\nu\nonumber \\
 & =\frac{\sin\left(2\pi^{2}\omega_{0}\left(\kappa-2\right)\right)}{16\pi^{2}\omega_{0}\left(\kappa-2\right)}+\frac{\sin\left(2\pi^{2}\omega_{0}\left(\kappa+2\right)\right)}{16\pi^{2}\omega_{0}\left(\kappa+2\right)}+\frac{\sin\left(2\pi^{2}\kappa\omega_{0}\right)}{8\pi^{2}\omega_{0}\left(\kappa+2\right)}
\end{aligned}
}%
\end{equation}%
\begin{equation}%
{
\begin{aligned}
\text{cov}\left(x\left(n-1\right),y\left(n-2\right)\right) & =\frac{1}{2\pi}\int_{0}^{2\pi}\left(\cos\left(2\pi\kappa\omega_{0}\nu\right)\cos\left(2\pi\omega_{0}\nu\right)\cos\left(2\pi\omega_{0}\left(\nu-1\right)\right)\right)\,d\nu\nonumber \\
 & =\frac{\sin\left(4\pi^{2}\omega_{0}\left(\kappa-2\right)+2\pi\omega_{0}\right)}{16\pi^{2}\omega_{0}\left(\kappa-2\right)}+\frac{\sin\left(4\pi^{2}\omega_{0}\left(\kappa+2\right)-2\pi\omega_{0}\right)}{16\pi^{2}\omega_{0}\left(\kappa+2\right)}\nonumber \\
 & +\frac{\sin\left(4\pi\omega_{0}\left(\pi\kappa+\frac{1}{2}\right)\right)}{16\pi^{2}\omega_{0}\kappa}+\frac{\sin\left(4\pi\omega_{0}\left(\pi\kappa-\frac{1}{2}\right)\right)}{16\pi^{2}\omega_{0}\kappa}\nonumber \\
 & +\frac{\sin\left(\pi\omega_{0}\right)\cos\left(\pi\omega_{0}\right)}{2\pi^{2}\omega_{0}\left(\kappa^{2}-4\right)}\end{aligned}
}%
\end{equation}%
\begin{equation}%
{
\begin{aligned}
\text{cov}\left(x\left(n-2\right),y\left(n\right)\right) & =\frac{1}{2\pi}\int_{0}^{2\pi}\left(\cos\left(2\pi\kappa\omega_{0}\left(\nu-2\right)\right)\cos\left(2\pi\omega_{0}\left(\nu-2\right)\right)\cos\left(2\pi\omega_{0}\left(\nu-1\right)\right)\right)\,d\nu\nonumber \\
 & =\frac{\sin\left(4\pi\omega_{0}\left(\pi\left(\kappa-2\right)-\left(\kappa-1\right)\right)\right)}{16\pi^{2}\omega_{0}\left(\kappa-2\right)}+\frac{\sin\left(4\pi\omega_{0}\left(\pi\left(\kappa+2\right)-\left(\kappa+1\right)\right)\right)}{16\pi^{2}\omega_{0}\left(\kappa+2\right)}\nonumber \\
 & +\frac{\sin\left(4\pi\omega_{0}\left(\pi\kappa+\left(\kappa-1\right)\right)\right)}{16\pi^{2}\omega_{0}}+\frac{\sin\left(4\pi\omega_{0}\left(\pi\kappa-\left(\kappa-1\right)\right)\right)}{16\pi^{2}\omega_{0}}\nonumber \\
 & +\frac{\sin\left(4\pi\omega_{0}\left(\kappa-1\right)\right)}{8\pi^{2}\omega_{0}\kappa\left(\kappa-2\right)}\left(\kappa-1\right)+\frac{\sin\left(4\pi\omega_{0}\left(\kappa+1\right)\right)}{8\pi^{2}\omega_{0}\kappa\left(\kappa+2\right)}\left(\kappa+1\right)
 \end{aligned}
}%
\end{equation}%
Assuming $\text{cov}\left(x\left(n-1\right),y\left(n-1\right)\right)=\text{cov}\left(x\left(n-1\right),y\left(n\right)\right)$

In \autoref{tab:covariance}, we collect some boundaries to denote the proximity of the values to zero. Let be $\kappa\ge20$, so we can approximate%
\begin{equation}%
{
\begin{aligned}
\left(\begin{array}{c}
\hat{\phi}_{x1}\\
\hat{\phi}_{x2}\\
\hat{\phi}_{y1}\\
\hat{\phi}_{y2}
\end{array}\right) & \approx\left(\begin{array}{cccc}
\gamma_{x}\left(0\right) & \gamma_{x}\left(1\right) & 0 & 0\\
\gamma_{x}\left(1\right) & \gamma_{x}\left(0\right) & 0 & 0\\
0 & 0 & \gamma_{y}\left(0\right) & \gamma_{y}\left(1\right)\\
0 & 0 & \gamma_{y}\left(1\right) & \gamma_{y}\left(0\right)
\end{array}\right)^{-1}\left(\begin{array}{c}
0\\
0\\
\gamma_{y}\left(1\right)\\
\gamma_{y}\left(2\right)
\end{array}\right)
\end{aligned}
}%
\end{equation}%
Recall that the inverse of a block diagonal matrix is a block matrix composed of the inverse matrices,%
\begin{equation}%
{
\begin{aligned}
\left(\begin{array}{c}
\hat{\phi}_{x1}\\
\hat{\phi}_{x2}
\end{array}\right) & \approx\left(\begin{array}{c}
0\\
0
\end{array}\right)\\
\left(\begin{array}{c}
\hat{\phi}_{y1}\\
\hat{\phi}_{y2}
\end{array}\right) & \approx\left(\begin{array}{cc}
\gamma_{y}\left(0\right) & \gamma_{y}\left(1\right)\\
\gamma_{y}\left(1\right) & \gamma_{y}\left(0\right)
\end{array}\right)^{-1}\left(\begin{array}{c}
\gamma_{y}\left(1\right)\\
\gamma_{y}\left(2\right)
\end{array}\right)
\end{aligned}
}%
\end{equation}%
Finally, a VAR(2) cannot show any linear dependency between $x\left(n\right)$ and $y\left(n\right)$.
\end{proof}

\section{Correlation and frequency}\label{sec:Appendix-Multicollinearity}

\begin{table}{%
\centering
\includegraphics[width=1\textwidth,height=\textheight]{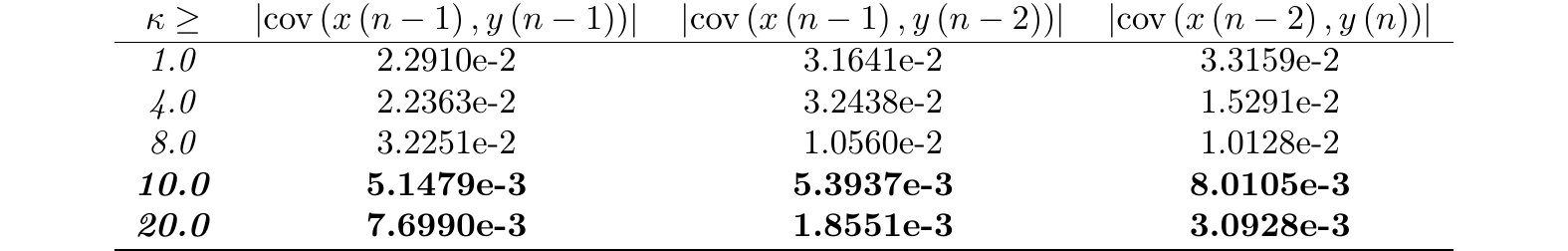}
\caption{Maximum covariance between $x$ and $y$ for lags 0, 1 and 2 in the interval $0.05\le\omega\le0.5$. Note that covariances are highly dependent on the amplifier magnitude $\kappa$. For $\kappa\ge10$, the three covariances are 4.14 to 5.87 times lower than the covariance at $\kappa=1$.}\label{tab:covariance}
}
\end{table}

\begin{lemma*}[]\label{thm:correlation-frequency0}Assume two zero-mean time series, $x_{A}\left(n\right)$ and $x_{B}\left(n\right)$, described through a second-order autoregressive model,%
\begin{equation}%
{
\begin{aligned}
x_{A}\left(n\right) & =\phi_{1A}x_{A}\left(n-1\right)+\phi_{2A}x_{A}\left(n-2\right)+\epsilon_{A}\left(n\right)\\
x_{B}\left(n\right) & =\phi_{1B}x_{B}\left(n-1\right)+\phi_{2B}x_{B}\left(n-2\right)+\epsilon_{B}\left(n\right)
\end{aligned}
}%
\end{equation}%
with correlated $\epsilon_{A}$ and $\epsilon_{B}$, resonating frequencies as $f_{A}^{*}$ and $f_{B}^{*}$, and recorded at a sampling frequency $f_{s}$. Both time series $x_{A}\left(n\right)$ and $x_{B}\left(n\right)$, present a cross-correlation at the first lag $\rho_{AB}\left(1\right)$ and $\rho_{BA}\left(1\right)$ given by%
\begin{equation}%
{
\begin{aligned}
\rho_{AB}\left(1\right) 
  & =2\rho_{AB}\left(0\right)
  \cos\left(2\pi\omega_{A}^{*}\right)
  \sqrt{-\phi_{2A}}
  \frac{1+\sqrt{\phi_{2A}\phi_{2B}}
          \frac{\cos\left(2\pi\omega_{B}^{*}\right)}
               {\cos\left(2\pi\omega_{A}^{*}\right)}
       }{1-\phi_{2A}\phi_{2B}}\\
\rho_{BA}\left(1\right)
  & =2\rho_{AB}\left(0\right)
 \cos\left(2\pi\omega_{B}^{*}\right)
 \sqrt{-\phi_{2B}}
 \frac{1+\sqrt{\phi_{2A}\phi_{2B}}
         \frac{\cos\left(2\pi\omega_{A}^{*}\right)}
         {\cos\left(2\pi\omega_{B}^{*}\right)}
      }{1-\phi_{2A}\phi_{2B}}
\end{aligned}
}%
\end{equation}%
where $\omega_{A}^{*}=\frac{f_{A}^{*}}{f_{s}}$ and $\omega_{B}^{*}=\frac{f_{B}^{*}}{f_{s}}$, $\rho_{AB}\left(0\right)$ is the correlation between $x_{A}$ and $x_{B}$.\end{lemma*}

\begin{lemma*}[Multicollinearity]\label{lem:multicollinearity0}Both zero-mean time series, $x_{A}\left(n\right)$ and $x_{B}\left(n\right)$ , with identical resonating frequencies $f^{*}\le0.02f_{s}$ with frequency bandwidths $\tau\ge3$ have a cross-correlation at the first lag $\rho_{AB}\left(1\right)>0.991$.\end{lemma*}

\begin{proof}Recall that coefficients $\phi_{1,k}$ and $\phi_{2,k}$ for $k=\left\{ A,B\right\}$ can be parametrized based on the central frequency $f_{k}^{*}=\omega_{k}^{*}f_{s}$ of the signal and the frequency bandwidth $\tau_{k}$:%
\begin{equation}%
{
\begin{aligned}
\phi_{1,k} & =\frac{2}{1+e^{-\tau_{k}}}\cos\left(2\pi\omega_{k}^{*}\right)\\
\phi_{2,k} & =-\frac{1}{\left(1+e^{-\tau_{k}}\right)^{2}}
\end{aligned}
}%
\end{equation}%
The bandwidth $\tau\rightarrow\infty$ generates a pure sinusoid and $\tau\rightarrow\infty$ resemble a low- or a high-pass filter according to the value of $f_{k}^{*}$.

\begin{table}{%
\centering
\includegraphics[width=1\textwidth,height=\textheight]{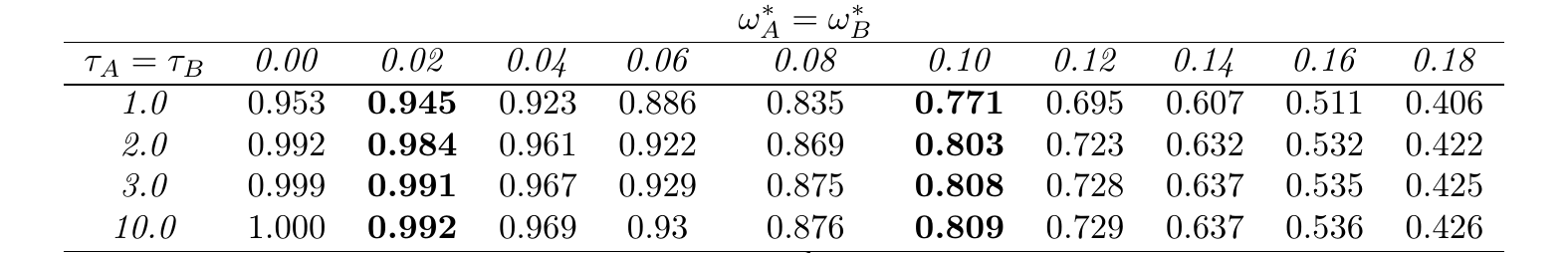}
\caption{Cross-correlation $\rho_{AB}\left(1\right)$ between two independent AR(2) time series $x_{A}\left(t\right)$ and $x_{B}\left(t\right)$ with same normalized central frequencies $\omega_{A}^{*}=\omega_{B}^{*}$ and frequency bandwidths $\tau_{A}=\tau_{B}$. The highlighted values can correspond to the 4Hz and 20Hz frequencies with a sampling frequency of 200Hz. Note that the correlation at 20Hz never reach a value close to 1, and therefore, can avoid multicollinearity issues during the estimation of parameters.}\label{tab:cross-correlation}
}
\end{table}

Note that the auto-correlation of $x_{A}\left(n\right)$%
\begin{equation}%
{
\begin{aligned}
\rho_{A}\left(1\right) & =\frac{\phi_{1}}{1-\phi_{2}}\nonumber \\
 & =2\cos\left(2\pi\omega^{*}\right)\sqrt{-\phi_{2}}\frac{1}{1-\phi_{2}}
\end{aligned}
}%
\end{equation}%

The covariance between $x_{A}\left(n\right)$ and $x_{B}\left(n-1\right)$ , given the imposed zero-mean condition, is described by%
\begin{equation}%
{
\begin{aligned}
E\left[x_{A}\left(n\right)x_{B}\left(n-1\right)\right] & =E\left[\left(\phi_{1A}x_{A}\left(n-1\right)+\phi_{2A}x_{A}\left(n-2\right)+\epsilon_{A}\left(n\right)\right)x_{B}\left(n-1\right)\right]\nonumber \\
 & =\phi_{1A}E\left[x_{A}\left(n-1\right)x_{B}\left(n-1\right)\right]+\phi_{2A}E\left[x_{A}\left(n-2\right)x_{B}\left(n-1\right)\right]\nonumber \\
 & =\phi_{1A}E\left[x_{A}\left(n\right)x_{B}\left(n\right)\right]+\phi_{2A}E\left[x_{A}\left(n-1\right)x_{B}\left(n\right)\right]
\end{aligned}
}\label{eq:recursion-1}%
\end{equation}%

Similarly, the covariance between $x_{A}\left(n-1\right)$ and $x_{B}\left(n\right)$ is%
\begin{equation}%
{
\begin{aligned}
E\left[x_{A}\left(n-1\right)x_{B}\left(n\right)\right] & =E\left[x_{A}\left(n-1\right)\left(\phi_{1B}x_{B}\left(n-1\right)+\phi_{2B}x_{B}\left(n-2\right)+\epsilon_{B}\left(n\right)\right)\right]\nonumber \\
 & =\phi_{1B}E\left[x_{A}\left(n-1\right)x_{B}\left(n-1\right)\right]+\phi_{2B}E\left[x_{A}\left(n-1\right)x_{B}\left(n-2\right)\right]\nonumber \\
 & =\phi_{1B}E\left[x_{A}\left(n\right)x_{B}\left(n\right)\right]+\phi_{2B}E\left[x_{A}\left(n\right)x_{B}\left(n-1\right)\right]
\end{aligned}
}\label{eq:recursion-2}%
\end{equation}%

We resolve the recursion replacing \autoref{eq:recursion-2} in \autoref{eq:recursion-1}:%
\begin{equation}%
{
\begin{aligned}
E\left[x_{A}\left(n\right)x_{B}\left(n-1\right)\right] & =E\left[\left(\phi_{1A}x_{A}\left(n-1\right)+\phi_{2A}x_{A}\left(n-2\right)+\epsilon_{A}\left(n\right)\right)x_{B}\left(n-1\right)\right]\nonumber \\
 & =\phi_{1A}E\left[x_{A}\left(n-1\right)x_{B}\left(n-1\right)\right]+\phi_{2A}E\left[x_{A}\left(n-2\right)x_{B}\left(n-1\right)\right]\nonumber \\
 & =\phi_{1A}E\left[x_{A}\left(n\right)x_{B}\left(n\right)\right]+\phi_{2A}\phi_{1B}E\left[x_{A}\left(n\right)x_{B}\left(n\right)\right]\nonumber \\
 & +\phi_{2A}\phi_{2B}E\left[x_{A}\left(n\right)x_{B}\left(n-1\right)\right]
\end{aligned}
}%
\end{equation}%

Simplifying, the correlation $\rho_{AB}\left(1\right)$ is%
\begin{equation}%
{
\begin{aligned}
\rho_{AB}\left(1\right)=\frac{E\left[x_{A}\left(n\right)x_{B}\left(n-1\right)\right]}{E\left[x_{A}\left(n\right)x_{B}\left(n\right)\right]} & \rho_{AB}\left(0\right)=\rho_{AB}\left(0\right)\phi_{1A}\frac{1+\phi_{2A}\frac{\phi_{1B}}{\phi_{1A}}}{1-\phi_{2A}\phi_{2B}}\nonumber \\
 & =2\rho_{AB}\left(0\right)\cos\left(2\pi\omega_{A}^{*}\right)\sqrt{-\phi_{2A}}\frac{1+\sqrt{\phi_{2A}\phi_{2B}}\frac{\cos\left(2\pi\omega_{B}^{*}\right)}{\cos\left(2\pi\omega_{A}^{*}\right)}}{1-\phi_{2A}\phi_{2B}}
\end{aligned}
}%
\end{equation}%

And, similarly, $\rho_{BA}\left(1\right)$:%
\begin{equation}%
{
\begin{aligned}
\rho_{BA}\left(1\right)=\frac{E\left[x_{A}\left(n-1\right)x_{B}\left(n\right)\right]}{E\left[x_{A}\left(n\right)x_{B}\left(n\right)\right]} & \rho_{AB}\left(0\right)=\rho_{AB}\left(0\right)\phi_{1B}\frac{1+\phi_{2B}\frac{\phi_{1A}}{\phi_{1B}}}{1-\phi_{2A}\phi_{2B}}\nonumber \\
 & =2\rho_{AB}\left(0\right)\cos\left(2\pi\omega_{B}^{*}\right)\sqrt{-\phi_{2B}}\frac{1+\sqrt{\phi_{2A}\phi_{2B}}\frac{\cos\left(2\pi\omega_{A}^{*}\right)}{\cos\left(2\pi\omega_{B}^{*}\right)}}{1-\phi_{2A}\phi_{2B}}
\end{aligned}
}%
\end{equation}%

Some numerical values for a set of $\tau_{k}$ and $\omega_{k}^{*}$ is shown in \autoref{tab:cross-correlation}, proving also the lemma.
\end{proof}

\bibliographystyle{elsarticle-num}
\bibliography{Library,ExtraBib}

\end{appendices}

\end{document}